\documentclass{article}
\usepackage[utf8]{inputenc}

\title{Providing early indication of regional anomalies in COVID19 case counts in England using search engine queries}

\author{Elad Yom-Tov\footnote{Microsoft Research, Herzeliya, Israel and Faculty of Industrial Engineering and Management, Technion, Israel}, Vasileios Lampos\footnote{Department of Computer Science, University College London, London, UK}, Ingemar J. Cox\footnote{Department of Computer Science, University College London, London, UK}\footnote{Department of Computer Science, University of Copenhagen, Denmark}, Michael Edelstein\footnote{National Infection Service, Public Health England, London, UK}}
\date{June 2020}

\usepackage{url}
\usepackage{graphicx}
\usepackage{subfigure}
\usepackage{enumerate}
\usepackage[shortlabels]{enumitem}

\begin{document}

\maketitle

\section*{Abstract}

COVID19 was first reported in England at the end of January 2020, and by mid-June over 150,000 cases were reported. We assume that, similarly to influenza-like illnesses, people who suffer from COVID19 may query for their symptoms prior to accessing the medical system (or in lieu of it). Therefore, we analyzed searches to Bing from users in England, identifying cases where unexpected rises in relevant symptom searches occurred at specific areas of the country.

Our analysis shows that searches for ``fever'' and ``cough'' were the most correlated with future case counts, with searches preceding case counts by 16-17 days. Unexpected rises in search patterns were predictive of future case counts multiplying by 2.5 or more within a week, reaching an Area Under Curve (AUC) of 0.64. Similar rises in mortality were predicted with an AUC of approximately 0.61 at a lead time of 3 weeks.

Thus, our metric provided Public Health England with an indication which could be used to plan the response to COVID19 and could possibly be utilized to detect regional anomalies of other pathogens.

\section{Introduction}

COVID19 was first reported in England in late January 2020~\cite{moss2020}. By the middle of June 2020 over 150,000 cases and 39,000 deaths were reported. 

In early March 2020, Public Health England (PHE), University College London (UCL) and Microsoft began investigating the possibility of using Bing search data to detect areas where outbreaks of the disease might be occurring or are soon to occur, so as to assist PHE in better planning their response.

Internet data in general and search data in particular, have long been used to track Influenza-Like Illness (ILI)~\cite{lampos2015gft,yang2015accurate,wagner2018}, norovirus~\cite{edelstein2014detecting}, and dengue fever~\cite{copeland2013} in the community. The main reason for the utility of these data for this purpose is the fact that most people with, for example, ILI will not visit a medical facility but will search about it or mention it in social media postings~\cite{yomtov2016crowdsourced}. We assume that the similarity of symptoms between ILI and COVID19, together with public fear of accessing medical facilities during an epidemic, may drive people to similarly search the web for relevant symptoms, making them predictive of COVID19.

Models of ILI which are based on internet data are usually trained using past season's data. Since this was infeasible for COVID19 we opted to use a different approach in our prediction, which utilized less training data. Our methodology examined consecutive weeks, where during the first of those weeks we found, for each Upper Tier Local Authority (UTLA), other UTLAs with similar rates of queries for symptoms. These UTLAs were then utilized to predict the rate of symptom queries for relevant symptoms during the following week. The difference between the actual and predicted rate of searches served as an indication of an unusual number of searches in a given area, i.e., an anomaly. 

This methodology is similar to a difference-in-difference analysis \cite{dimick2014}, albeit one where differences are calculated between actual and predicted symptom rates. As such, it shares similarities with the methodology used to predict the effectiveness of childhood flu vaccinations using internet data~\cite{lampos2015assessing,wagner2017}.

\section{Methods}

\subsection{Symptom list and area list}

The list of 25 relevant symptoms for COVID19 was extracted from PHE reports, and are listed in Table~\ref{tab:keywords} together with their synonyms.

In order to maximise the utility of the analysis, we conducted it at the level of UTLA, a subnational administrative division of England into 173 areas, over which local government has a public health remit.

\begin{table}[tp]
\centering
\begin{tabular}{|l|l|} 
\hline
\hline
COVID19 symptoms & Synonyms or related expressions\\
\hline
Altered consciousness & altered consciousness \\ 
Anorexia & appetite loss, loss of appetite, lost appetite \\ 
Anosmia & loss of smell, can't smell \\
Arthralgia & joint ache, joint aching, joints ache, joints aching \\
Chest pain & chest pain \\
Chills & chills \\
Cough & cough \\
Diarrhea & diarrhea, diarrhoea \\
Dry cough & dry cough \\
Dyspnea & breathing difficult, short breath, shortness of breath \\
Epistaxis & nose bleed, nose bleeding \\
Fatigue & fatigue \\
Head ache & head ache, headache \\
Myalgia & muscle ache, muscular pain \\
Nasal congestion & blocked nose, nasal congestion \\
Nausea & nausea, nauseous \\
Pyrexia & fever, high temperature \\
Pneumonia & pneumonia, respiratory infection, respiratory symptoms \\
Rash & rash \\
Rhinorrhea & runny nose \\ 
Seizure & seizure \\
Sore throat & sore throat, throat pain \\ 
Sternutation & sneeze, sneezing \\
Tiredness & tiredness \\
Vomiting & vomit, vomiting \\
\hline
\hline
\end{tabular}
  \caption{25 symptoms related to COVID19 (as identified by PHE) and their synonyms or related expressions.}
  \label{tab:keywords}
\end{table}

\subsection{Search data}

We extracted all queries submitted to the Bing search engine from users in England. Each query was mapped to a UTLA according to the postcode (derived from the IP address of the user) from which the user was querying. We counted the number of users per week who queried for each of the keywords from each UTLA, and normalized by the number of users who queried for any topic during that week from each UTLA. The fraction of users who queried for keyword $k$ at week $w$ in UTLA $i$ is denoted by $F_{wk}^i$.

Data was extracted for the period between January 1st, 2020 to May 28th, 2020. For privacy reasons, UTLAs with fewer than 10,000 Bing users were removed from the analysis. Additionally, any keyword which had fewer than 10 users in a given week at a specific UTLA was also removed from the analysis by setting it to zero.

This study was approved by the Institutional Review Board of the Technion, Israel Institute of Technology.

\subsection{Validation data}

We compared our detection methodology (described below) to mortality data and case reports. The former were obtained at a weekly resolution from the UK Office of National Statistics Death Registrations and Occurrences by Local Authority and Health Board\footnote{\url{https://www.ons.gov.uk/peoplepopulationandcommunity/healthandsocialcare/causesofdeath/datasets/deathregistrationsandoccurrencesbylocalauthorityandhealthboard}}. Case report numbers per day were accessed from UK government's dashboard for COVID19\footnote{\url{https://coronavirus.data.gov.uk/}}.

\subsection{Analysis}

Analysis was conducted at a weekly resolution, beginning on Mondays of each week, starting on March 4th, 2020. For a given week $w$ we found control UTLAs for each UTLA such that $F_{wk}^i$ could be predicted from $F_{wk}^j$ ($j=1,2,...,5$). To do this, a greedy procedure was followed for each UTLA $i$:

\begin{enumerate}[start=1,label={\bfseries(\arabic*)}]
    \item Find a UTLA which is at least 50km distant from the $i$-th UTLA for which the linear function $F_{wk}^i=f(F_{wk}^j)$ mapping the symptom rates at $j$ to the symptom rates at $i$ reaches the highest coefficient of determination ($R^2$). That is, for each $k$, $F_{wk}^i \approx f(F_{wk}^j)$ in a least-squares sense.
    
    \item Repeat \textbf{(1)}, adding at each time another area that maximally increases $R^2$ when added to the previously established set of areas. 
\end{enumerate}

The linear function $f$ was optimized for a least squares fit, with an intercept term.

The result of this procedure is a linear function which predicts the symptom rate for each UTLA given the symptom rates at 5 other UTLAs at week $w$. We denote this prediction as 
$ \hat{F}_{wk}^i = f(F_{wk}^j)$.

The function is applied at week $(w+1)$ to each UTLA, and the difference between the estimated and actual symptom rate for each symptom is calculated: $d_k^i = F_{(w+1)k}^i - \hat{F}_{(w+1)k}^i$. We refer to this difference as the {\bf UTLA outlier measure}.

To facilitate comparison between the differences of different keywords, $d_k^i$, the values of $d_k^i$ are normalized to zero mean and unit variance (standardized) for each keyword.

\section{Results}

\subsection{Correlation of individual keywords with case counts}

\begin{figure}
\begin{center}
  \includegraphics[width=0.8\linewidth]{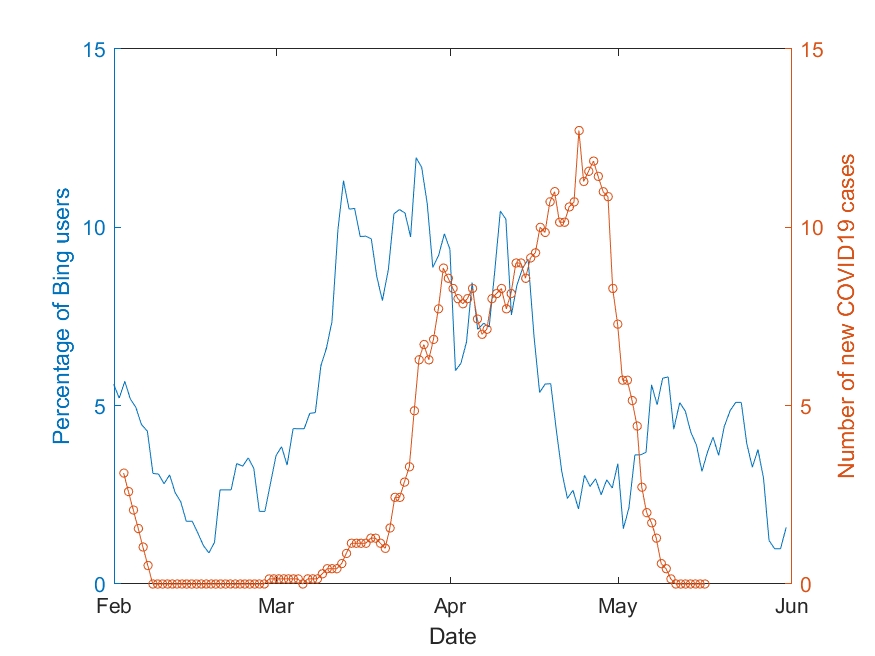}
\end{center}
  \caption{Number of COVID19 cases (red circles) and percentage of Bing users who queried for ``cough'' in a sample UTLA (blue). Curves are smoothed using a moving average filter of length 7. The correlation between the curves is 0.837 at a delay of 20 days.}
  \label{fig:sample_corr}
\end{figure}

For illustrative purposes, Figure~\ref{fig:sample_corr} shows the daily number of COVID19 cases and percentage of Bing users who queried for ``cough'' in one of the UTLAs. 
We calculated the cross-correlation between these time series for each keyword and  each UTLA. The highest correlation and its lag in days were noted, and the median values (across UTLAs) are shown for each keyword in Table~\ref{tab:corr_keywords}. 

As the Table shows, the best correlations are reached for cough, sore throat, and fever, at a delay of 16-19 days. Based on initial results and using PHE case definition of COVID19 at the time, we focused on two keywords, cough and fever, for the the remaining analysis.

\begin{table}[tp]
\centering
\begin{tabular}{|l|c|c|} 
\hline
\hline
Keyword & Median correlation & Median lag (days) \\
\hline
chest pain & 0.589 &   13 \\
cough & 0.746 &   17 \\
diarrhea & 0.606 &   22 \\
fatigue & 0.509 &  -13 \\
fever & 0.695 &   16 \\
head ache & 0.624 &   13 \\
nausea & 0.590 &   -4 \\
pneumonia & 0.667 &   34 \\
rash & 0.612 &   -8 \\
seizure & 0.579 &    6 \\
sneezing & 0.593 &    4 \\
sore throat & 0.775 &   19 \\
vomiting & 0.575 &   15 \\
\hline
\hline
\end{tabular}
  \caption{Median correlation and the lag (in days) at which it is achieved, between case numbers and fraction of keywords used on Bing. A positive lag means that Bing searches appear before case counts, and vice cersa. }
  \label{tab:corr_keywords}
\end{table}

\begin{figure}
\begin{center}
  \includegraphics[width=0.8\linewidth]{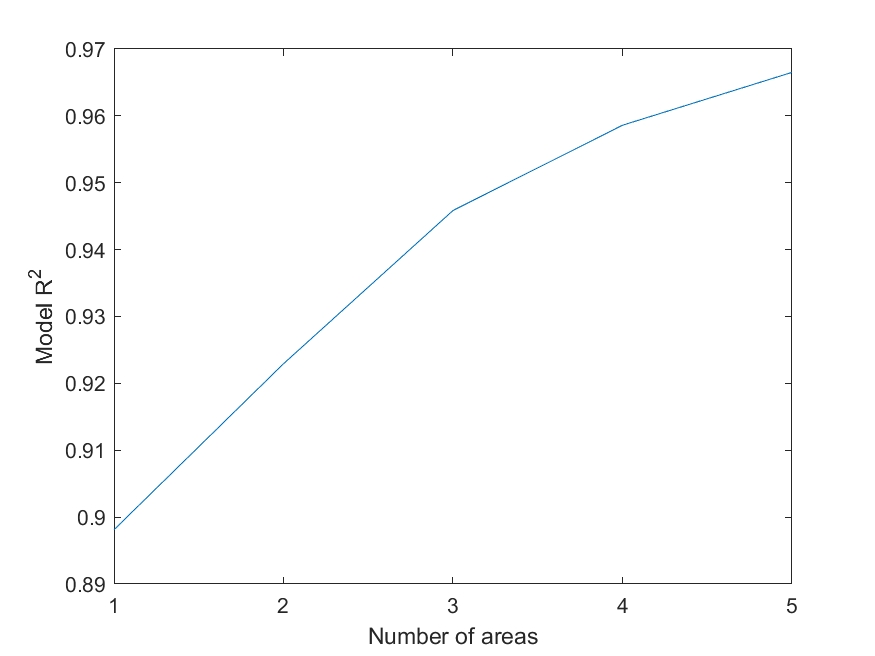}
\end{center}
  \caption{Average $R^2$ values of the model for $F_{wk}^i$ as the number of areas increases.}
  \label{fig:sample_r2}
\end{figure}

Figure~\ref{fig:sample_r2} shows the improvement in model fit ($R^2$) as more areas are added to $F_{wk}^i$. As the model shows, improvement continues, but the marginal gain decreases with the number of areas, as expected.

\subsection{Detection ability of the outlier measure} 

On average, predictions were given for 116 UTLAs per week (of 173 UTLAs), where at least 10,000 users queried on Bing.

\begin{figure}
\begin{center}
  \includegraphics[width=0.8\linewidth]{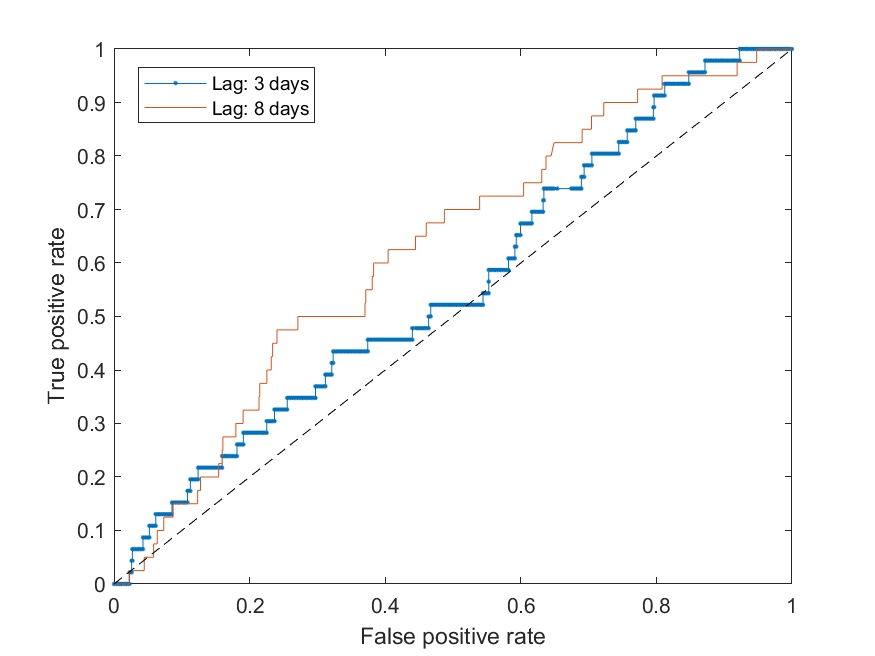}
\end{center}
  \caption{ROC curve for of the composite measure, derived from the product of the normalized UTLA outlier measures for ``fever'' and ``cough'', for two time lags (3 and 8 days).}
  \label{fig:auc_sample}
\end{figure}

Figure~\ref{fig:auc_sample} shows the ROC for two lags, 3 days (AUC: 0.56) and 8 days (AUC: 0.63) for the composite signal, that is, the multiplication of the normalized UTLA outlier measures for ``cough'' and ``fever''.

\begin{figure}
\begin{center}
  \includegraphics[width=0.8\linewidth]{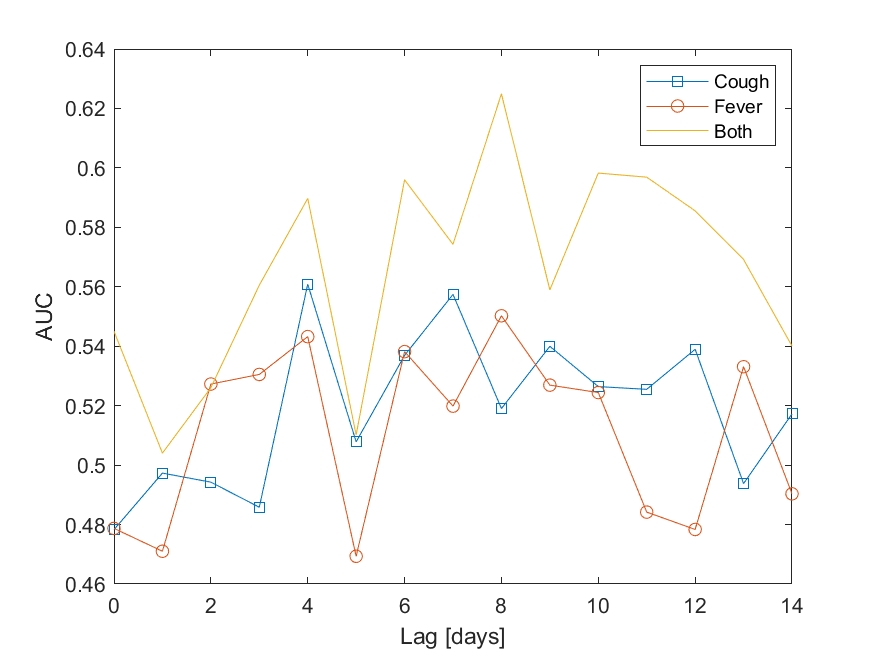}
\end{center}
  \caption{AUC of the UTLA outlier measure for detecting unusually large rises in COVID19 cases per UTLA, as a function the the lag between case counts and Bing data.}
  \label{fig:auc_over_time_cases}
\end{figure}

Figure~\ref{fig:auc_over_time_cases} shows the Area Under Curve (AUC) for the Receiver Operating Curve where the independent attribute is the UTLA outlier measures and the dependent variable is whether there was week-over-week jump of over two standard deviations in the number of COVID19 cases in an UTLA. As Figure~\ref{fig:auc_over_time_cases} shows, ``fever'' reaches a slightly higher AUC than ``cough'', but precedes case numbers by only around 2 days, meaning that it can predict cases with only limited lead time. In contrast, ``cough'' reaches a lower AUC at a lead time of 4-6 days. The multiplication of the values of both symptoms (denoted in the figure as ``Both'') reaches the highest AUC, at a lead of one week.

\begin{figure}
\begin{center}
  \includegraphics[width=0.8\linewidth]{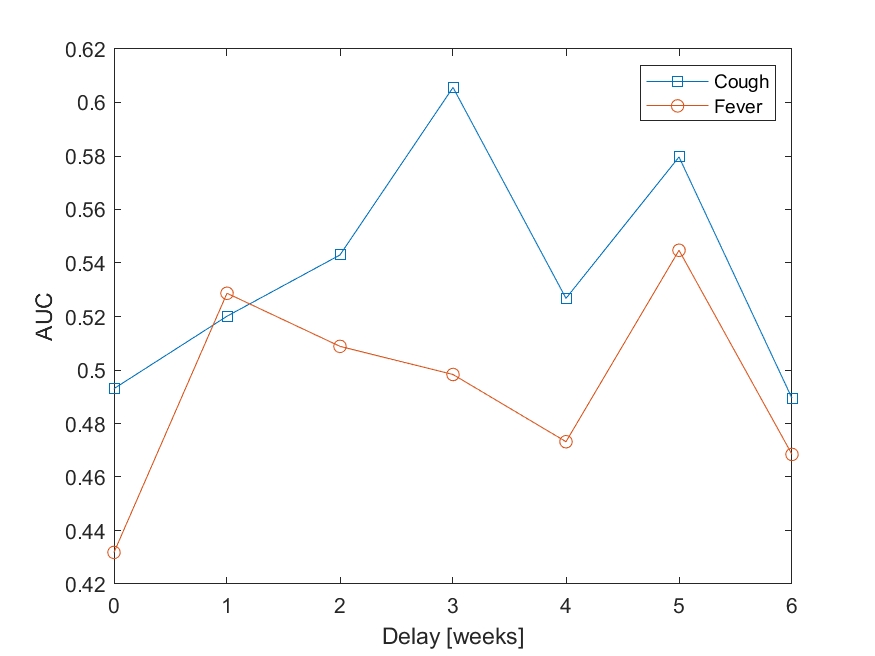}
\end{center}
  \caption{AUC of the UTLA outlier measure for detecting unusually large rises in COVID19 mortality per UTLA, as a function the the lag (in weeks) between mortality and Bing data. }
  \label{fig:auc_over_time_mortality}
\end{figure}

Figure~\ref{fig:auc_over_time_mortality} shows a similar analysis to the one above, albeit for mortality data at a weekly basis. As the Figure shows, cough is a better predictor, at a delay of 3 weeks. 

\subsection{Changes in detections over time} 

\begin{figure}
\begin{center}
\begin{tabular}{c}
\setlength{\tabcolsep}{0pt}
\renewcommand{\arraystretch}{0}
\includegraphics[width=0.6\linewidth]{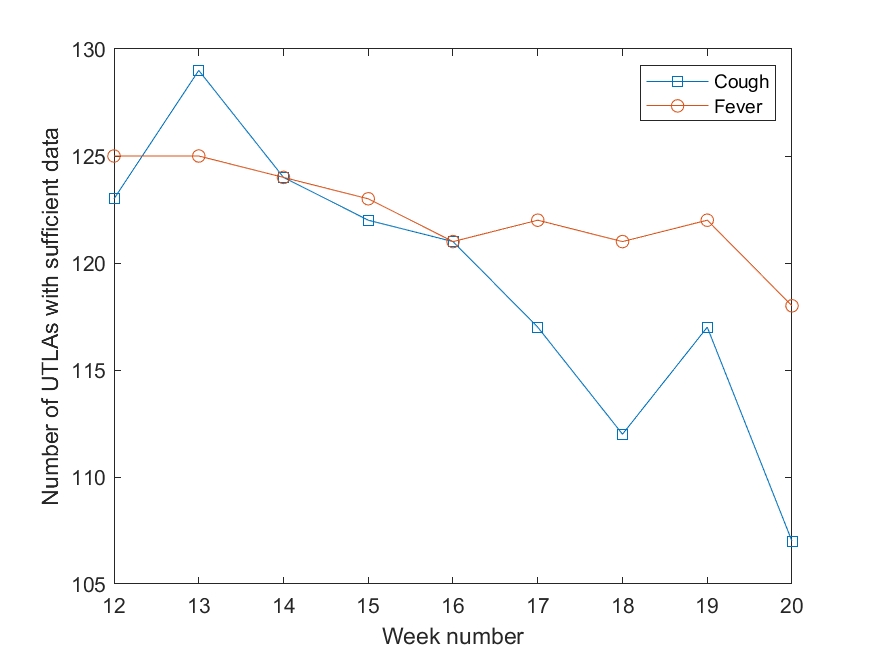}
\\
\includegraphics[width=0.6\linewidth]{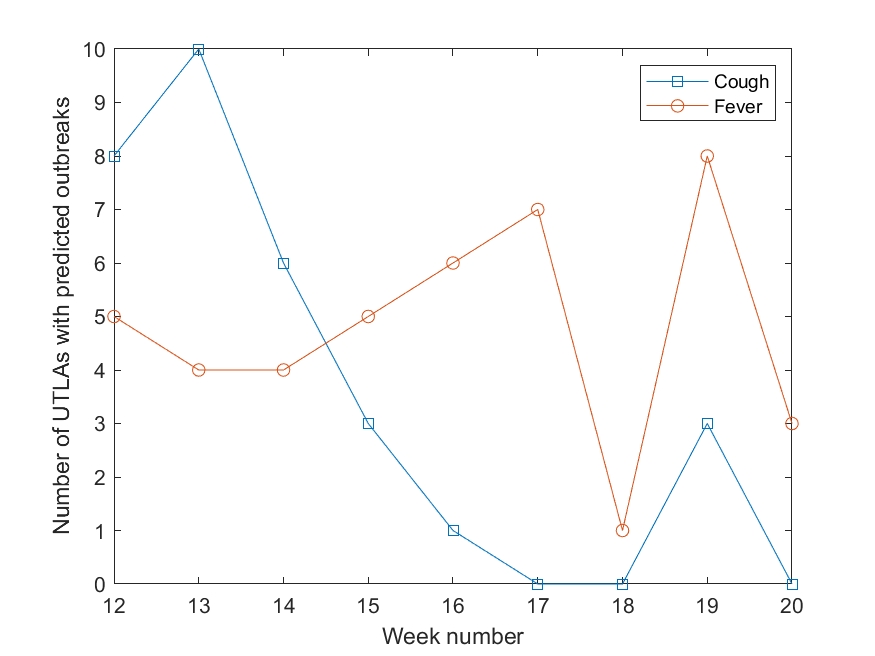} \\
\includegraphics[width=0.6\linewidth]{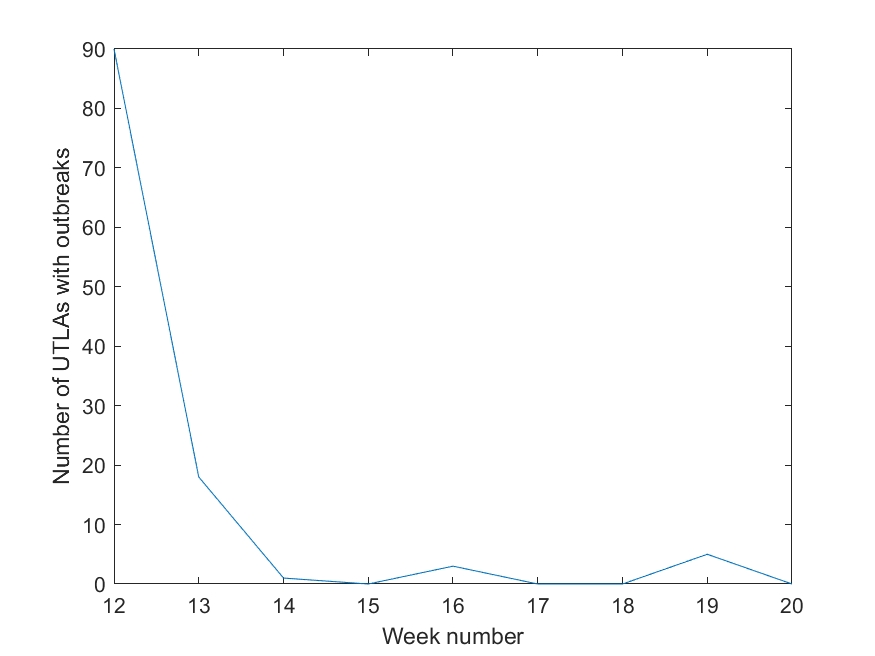} \\
\end{tabular}
\caption{Number of UTLAs with sufficient data over time (left), number of UTLAs with values over the threshold over time (middle) and number of UTLAs with rises greater than 2.5 times (right). Week numbers correspond to the weeks since the beginning of 2020.}
\label{fig:over_time}
\end{center}
\end{figure}

Figure~\ref{fig:over_time} shows the number of UTLAs with sufficient data, meaning that enough users queried for the relevant terms, over the weeks of the analysis. As the figure shows, the number of users asking about fever was relatively high throughout the analysis, but questions for cough dropped sufficiently that the number of UTLAs for which detections could be given was reduced by around a fifth. Figure~\ref{fig:over_time} (center) shows the number of UTLAs per week that had values greater than 2 standard deviations. Here too cough decreases quickly, while fever remains at a relatively constant level. Finally, the right figure shows the number of UTLAs which experienced a rise of 2.5 or more in the number of cases, week over week. Here too the number drops significantly over time.

\subsection{Demographic attributes of outlying areas}

The 10 highest and 10 lowest correct and mistaken detections at each week were identified to assess if they could by associated with specific demographic characteristics of their areas. 

Demographic characteristics of UTLAs were collected from the UK Office of National Statistics (ONS), and include population density, male and female life expectancy and healthy life expectancy, male to female ratio, and the percentage of the population under the age of 15. 

Association was estimated using a logistic regression model. However, none of the variables were statistically significantly associated with these attributes ($P>0.05$ with Bonferroni correction).

\section{Discussion}

Internet data, especially search engine queries, have been used for tracking influenza-like illness and other illnesses for over a decade, because of the frequency at which people query for the symptoms of these illnesses and the fact that more people search for them than visit a health provider~\cite{lampos2015gft,yang2015accurate}. COVID19, a novel disease, seemed to present similar opportunities for tracking using web data, and current indications suggest that search data could be used to track the disease~\cite{lampos2020}. However, as there was little past information to enable model training, we developed a method for detecting outbreaks using a variant of a difference-in-difference model at the local level. 

Our results demonstrate good correlation between case numbers and the use of the keywords ``cough'', ``fever'' and ``sore throat'', with queries leading case numbers by 16-19 days (similar to the findings of Lampos et al.~\cite{lampos2020}). Based on early indications from PHE we focused on using the first two keywords in our detection methodology. 

The detections provided to PHE provided a lead time of approximately one week for case numbers, with an AUC or approximately 0.64. This modest accuracy is nonetheless useful as long as exceedance of the 2 standard deviations threshold is not interpreted at face value as an increase in disease incidence, but as an early warning signal that triggers further investigation and correlation with outputs from other disease surveillance systems. This is particularly true at the local level and the outputs of this analysis are being incorporated into local routine PHE surveillance reports alongside outputs from clinical and laboratory systems. We also note that the detection accuracy for mortality was greatest at 3 weeks, which is congruent with the time difference between illness onset and death~\cite{Zhou2020}.

The threshold at which a UTLA should be alerted can be set in a number of ways. In our work with PHE, we reported UTLAs where the value of both symptoms multiplied exceeded the 95-th percentile threshold of values, computed for all other symptoms that week, similar to the procedure used in the False Detection Ratio test \cite{benjamini1995}.

The reasons for the modest detection accuracy include problems in the source data as well as in the data used as ground truth. Search data is noisy~\cite{yomtov2015} and Bing's market share in England is estimated at around 5\%~\cite{bingMarketShare}. We compared our results to the number of positive COVID19 cases. These numbers are affected by testing policy, which may have caused a non-uniform difference between known and actual case numbers in different UTLAs. Additionally, COVID19 has a relatively high asymptomatic rate (currently estimated at 40-45\%~\cite{covidasymptomatic}). People who do not experience symptoms would be missed by our method. On the other hand, current serological surveys~\cite{EnglandSerological} suggest that at the end of May 2020, between 5\% and 17\% of the population (depending on area in England) have been exposed to COVID19, compared to only 0.3\% than have tested positive to a screening test, suggesting that a large number of people who may have experienced symptoms of COVID19 and queried for them were not later tested, leading to errors in our comparison between detections and known case numbers.   

\bibliographystyle{plain}
\bibliography{references}

\end{document}